\documentclass{article}

\usepackage{PRIMEarxiv}

\usepackage[utf8]{inputenc} 
\usepackage[T1]{fontenc}    
\usepackage{hyperref}       
\usepackage{url}            
\usepackage{booktabs}       
\usepackage{amsfonts}       
\usepackage{nicefrac}       
\usepackage{microtype}      
\usepackage{lipsum}
\usepackage{fancyhdr}       
\usepackage{graphicx}       
\usepackage{amsmath}
\usepackage{threeparttable}
\usepackage{float}
\usepackage{booktabs}
\usepackage{graphicx}
\usepackage{booktabs}
\usepackage{multirow}
\graphicspath{{media/}}     

\title{Supervised Contrastive Learning for Recommendation
}

\author{
  Chun Yang \\
  School of Automation Engineering, University of Electronic Science and Technology of China \\
  Chengdu, Sichuan, China\\
  \texttt{beiluo@std.uestc.edu.cn} \\
}

\begin{document}
\maketitle

\begin{abstract}
Compared with the traditional collaborative filtering methods, the graph convolution network can explicitly model the interaction between the nodes of the user-item bipartite graph and effectively use higher-order neighbor information, which enables the graph neural network to obtain more effective embeddings for recommendation, such as NGCF And LightGCN. However, its representations are very susceptible to the noise of interaction. In response to this problem, SGL explored the self-supervised learning on the user-item graph to improve the robustness of GCN. Although effective, we found that SGL directly applies SimCLR's comparative learning framework. This framework may not be directly applicable to the scenario of the recommendation system, and does not fully consider the uncertainty of user-item interaction.

In this work, we aim to consider the application of contrastive learning in the scenario of the recommendation system adequately, making it more suitable for recommendation task. We propose a learning paradigm called supervised contrastive learning(SCL) to support the graph convolutional neural network. Specifically, we will calculate the similarity between different nodes in user side and item side respectively during data preprocessing, and then when applying contrastive learning, not only will the augmented views be regarded as the positive samples, but also a certain number of similar samples will be regarded as the positive samples, which is different with SimCLR that treats other samples in a batch as negative samples. We apply SCL on the most advanced LightGCN. In addition, in order to consider the uncertainty of node interaction, we also propose a new data augment method called node replication. Empirical research and ablation study on Gowalla, Yelp2018, Amazon-Book datasets prove the effectiveness of SCL and node replication, which improve the accuracy of recommendations and robustness to interactive noise.

\end{abstract}


\section{INTRODUCTION}
In order to meet the diverse and personalized information needs of users, the personalized recommendation system has emerged \cite{covington2016deep,zhou2019deep,pi2019practice,grbovic2018real}. A recommendation system can help users find items of interest in a large amount of candidate items. Early recommendation system which called collaborative filtering recommends items to users based on interactive content of similar users such as purchases, clicks, and ratings. Recent collaborative filtering system generally uses embeddings to represent users and items, and predicts scores based on corresponding embeddings. In recent years, a large number of collaborative filtering methods have been successfully implemented, such as \cite{koren2008factorization,he2017neural,ebesu2018collaborative,liang2018variational,hernando2016non}.


In recent years, in order to make use of the user-item bipartite graph high-hop neighbor information fully, collaborative filtering methods based graph convolution network(GCN) have been widely studied \cite{ying2018graph,zhang2019inductive,berg2017graph}. Representatively, Wang et al. recently proposed Neural Graph Collaborative Filtering (NGCF), and the best experimental results were obtained on three data sets \cite{wang2019neural}. They carried out feature embedding, information dissemination and feature aggregation on the graph to obtain the final representation of users and items. Based on NGCF, He et al. proposed lightGCN \cite{he2020lightgcn}. They believe that many operations of NGCF are directly inherited from GCN. These operations are not necessarily suitable for collaborative filtering tasks in the recommendation field. Therefore, they simplify many operations of NGCF, and achieve the current state-of-the-art on multiple data sets.


GCN provides a most advanced and efficient solution for solving user-item bipartite graph sparsity problems and interactive modeling problems. However, the representations learned by GCN are easily biased towards high-frequency items or users and are also easily affected by noise of interactions. In order to address the above limitations, Wu et al. proposed Self-supervised Graph Learning(SGL) for recommendation, which uses unlabeled data space by augmenting the input data, thereby achieving a significant improvement in downstream tasks \cite{wu2021self}.

Although SGL has achieved a certain degree of effect improvement, we believe that the contrastive learning framework of SGL may not be very suitable for recommendation tasks. Specifically, SGL use the contrastive learning framework of simCLR as a paradigm of self-supervised learning \cite{chen2020simple}, which treats augmenting samples of a input data as the positive samples and treats other samples in a same batch as the negative samples. Due to the diversity of samples in computer vision tasks, this design is reasonable and helps to mine hard negative samples which improve the quality of the learned representations. However, in the recommendation system, the target of the comparison is the user or item node, which means that there is a high probability that there are similar users or items in a batch. Regarding these samples as the negative samples, similar users or items will become farther away in the representation space, which violates the optimization purpose of the recommendation system, affects the final representation learned by GCN, and reduces the performance of the recommendation system.

\begin{figure*}
	\centering
		\includegraphics[scale=.48]{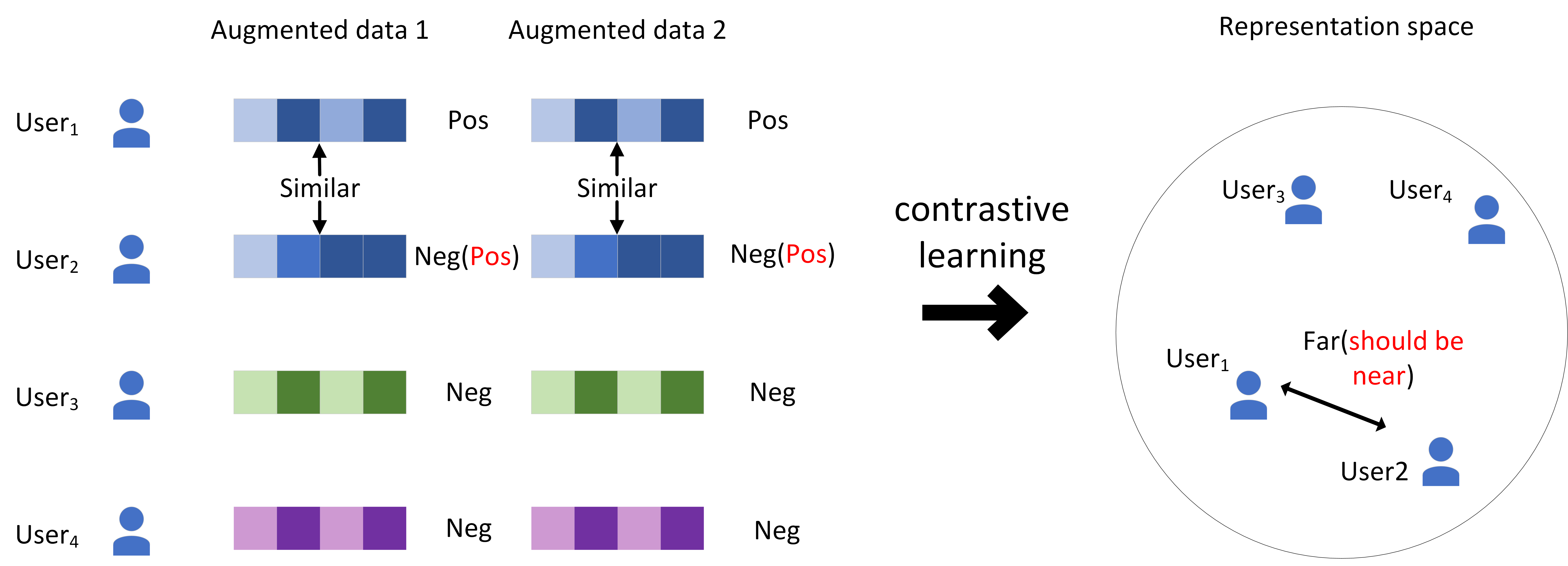}
	\caption{A toy example to show the limitation of contrastive learning framework used by SGL. SGL treats other samples in a batch as negative samples, which makes the representation of similar users further in the representation space, and it should be closer to be more effective for subsequent recommendation task. Our proposed SCL introduces supervised information and labels similar nodes as positive samples, as shown by the red word in the figure. After contrastive learning, similar nodes are closer in the representation space, which can obtain representations that are more conducive to recommendation.}
	\label{limitation}
\end{figure*}

In order to address the above limitation, we believe that it is necessary to improve the application of contrastive learning. We propose a learning paradigm called supervised contrastive learning(SCL) for recommendation. We design the contrastive learning framework guided by the recommendation task, so as to obtain the representations that is more in line with the requirements of the recommendation task and improve the performance of downstream tasks. A toy example of SCL is shown as figure \ref{limitation} . Specifically, it consists of two steps: (1) data augmentation, which generates multiple views for each node; (2) contrastive learning, which makes similar samples are closer and dissimilar samples are further in the representation space. We will not simply treat other samples in the same batch as negative samples when we perform comparative learning, but treat all similar samples as positive samples, and dissimilar samples as negative samples. This comparative learning method introduces supervised information, which makes similar users or items more inclined to learn similar representations, which is beneficial to downstream recommendation tasks. In this learning paradigm, we use multi-task learning to optimize model parameters, with the Bayesian Personalized Ranking(BPR) loss as the main task and contrastive learning loss as the auxiliary task \cite{rendle2012bpr}.

\begin{figure*}
	\centering
		\includegraphics[scale=.32]{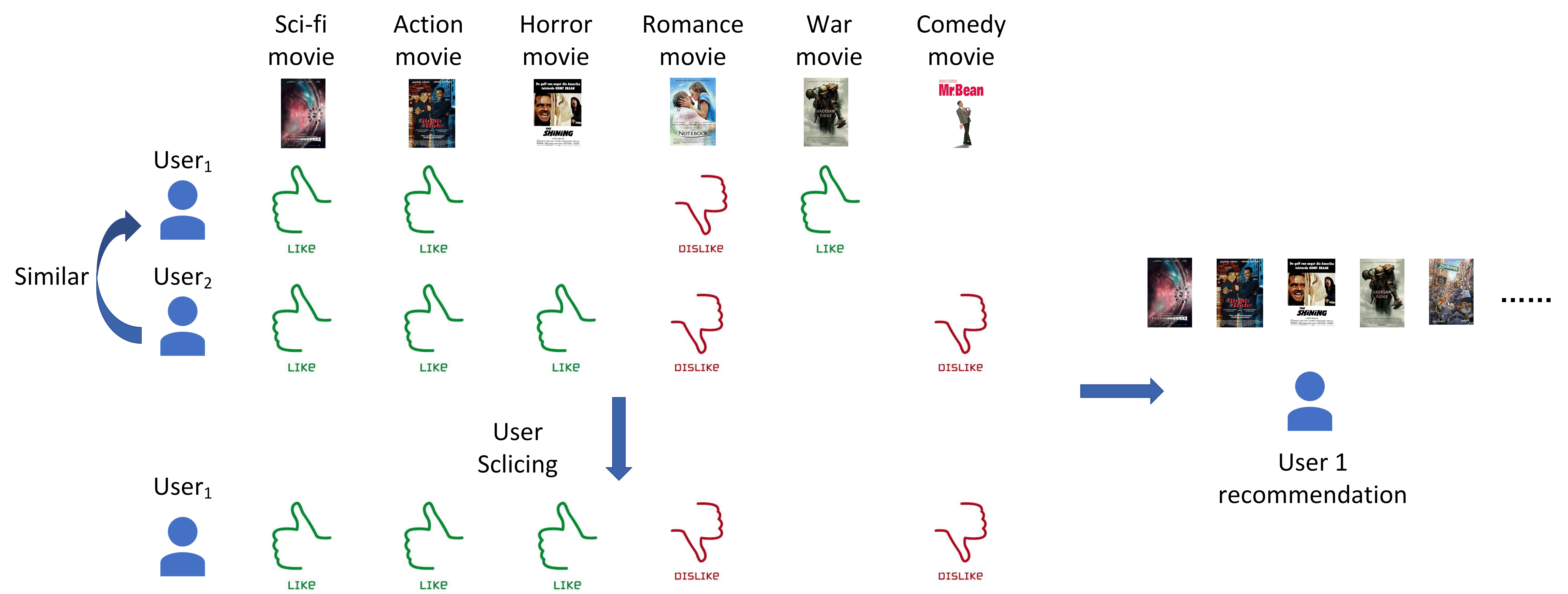}
	\caption{Motivation for node replication. By replacing part of the data of user 1 with the corresponding data of similar user 2, the augmented sample are more diverse, and the obtained representation can better express the user's preferences, thereby making the recommendation results more diverse.}
	\label{toy_example_user}
\end{figure*}

In addition, we found that the BPR loss which used in GCN is simply optimized by constructing interactive and non-interactive triples. However, the user-item bipartite graph may have noisy interaction or similar user-item pairs without interaction. We believe that the edge drop and node drop on the user-item bipartite graph does not bring enough diversity information. In order to improve the model's adaptability to noise interaction and the diversity of the representations, we propose a new data augmentation method called node replication(NR). Specifically, as shown in figure \ref{toy_example_user}, we replace part of the interaction of the current node with the corresponding interaction of similar nodes according to a certain probability. This data augmentation method utilizes supervised information, which can effectively improve recommendation performance and help increase the diversity of recommendations.

It is worth noting that our SCL can be applied to any collaborative filtering network based on the user-item bipartite graph. In this work, we choose to implement it on the most advanced LightGCN, and prove the effectiveness of SCL on three benchmark data set, which can significantly improve the recommendation performance.

To summarize, The main contributions of this paper are summarized as follows:
\begin{itemize}
\item We propose a new supervised comparative learning paradigm, which considers the purpose of the recommendation system and provides supervised information for representation learning, so that similar nodes are closer in the representation space, which helps to improve the recommendation performance.
\item In order to improve the representation ability of nodes and obtain more diverse information, we propose a new data augmentation method called node replication, which improves the robustness and diversity of the representations, and makes the recommendation results more diverse and has better performance.
\item We verified the effect of SCL on Gowalla, Yelp2018, Amazon-Book dataset and proved the effectiveness of SCL.
\end{itemize}


\section{PRELIMINARIES }
\label{PRELIMINARIES}

\subsection{PROBLEM STATEMENT}
\label{PROBLEM}

The set of user-item interactions can be easily modeled as a bipartite graph $ {\cal G} = ({\cal U},{\cal V},{\cal E})\ $, where $ {\cal U} $ represents the user set, $ {\cal V} $ represents the item set, and $ {\cal E} $ represents the edge set. If there is an interaction between a user and a item, then there is an edge between them \cite{sun2020neighbor}. The collaborative filtering based on GCN needs to embed each user $ i \in {\cal U} $ and item $ j \in {\cal V} $ into a unified d-dimensional representation space, and the corresponding representations are expressed by $ u_i \in {R^d} $ and $ v_j \in {R^d}$. By calculating the inner product between the representations of a user and a item we can know whether the item should be recommended to the user.

\subsection{GRAPH CONVOLUTIONAL NETWORK}
\label{GCN}

The core of GCN is to aggregate the domain representations of each node on the bipartite graph $ {\cal G} $, and update the representation of each node by considering the information of high-order neighbors to obtain an effective representation finally \cite{ying2018graph,zhang2019inductive,berg2017graph,wang2019neural,he2020lightgcn}. For user $ i $, its representation is calculated as follows:
\begin{equation}
\begin{aligned}
& {u_i}^{(l)} = \delta ({W^{(l)}}[{u^{\left( {l - 1} \right)}}_i|A({v_j} \in {\cal N}\left( {{u_i}} \right)])
\end{aligned}
\end{equation}

Where $ {v_j} \in {\cal N}\left( {{u_i}} \right) $ represents the neighbor domain nodes of user $ i $, and $ A $ denotes the node aggregation function, which can be average, weighted sum, linear mapping, etc \cite{gilmer2017neural,hamilton2017inductive,velivckovic2017graph}. $ [ \cdot \mid  \cdot ] $ is a concatenation operation and $ {W^{(l)}} $ is the weight matrix of the $ l $-th layer. $ \delta  $ is the activation function of the $ l $-th layer, which can be $ sigmoid $, $ Relu $, or None, etc \cite{xu2018powerful,hamilton2017inductive,zhu2020bilinear}. $ {u_i}^{(l)} $ denotes the representation of user $ i $ at layer $ l $, $ {u^{\left( {l - 1} \right)}}_i $ is that of previous layer. And $ {u^{\left( {0} \right)}}_i $ is the ID embeddings which are trainble parameters.

Specifically, GCN first aggregates the representations of all neighbor domain nodes of user $ i $ at layer $ l-1 $, then combines it with its own representation, and then obtains the representation of layer $ l $ through linear linear mapping. For a certain item $ j $, the calculation method of its representation is the same as that on the user side as follows:
\begin{equation}
\begin{aligned}
& {v_j}^{(l)} = \delta ({W^{(l)}}[{v^{\left( {l - 1} \right)}}_j|A({u_i} \in {\cal N}\left( {{v_j}} \right)])
\end{aligned}
\end{equation}

The obtained representation of the $ l $ layer corresponds to the information aggregation of the node's $ l $ hop neighbor. For different layers of information, there is a readout function to generate the final node representation, take the user side as an example:
\begin{equation}
\begin{aligned}
& {u_i} = {f_{{\rm{readout }}}}\left( {\left\{ {u_i^{(l)}\mid l = [0, \cdots ,L]} \right\}} \right)
\end{aligned}
\end{equation}

Where $ {f_{{\rm{readout }}}} $ is the readout function which can be the last layer only, splicing, averaging, weighted sum, attention and so on \cite{wang2019neural,he2020lightgcn,kipf2016semi,chen2017attentive,li2018deeper}. The calculation method on the item side is the same as that on the user side.

After obtaining the final representation of each node, the prediction layer can be used to calculate a certain user $ i $'s preference for item $ j $. In order to ensure the calculation efficiency of the GCN, the vector inner product is widely used as the prediction layer at present \cite{covington2016deep,grbovic2018real,he2020lightgcn,huang2020embedding,barkan2016item2vec,perozzi2014deepwalk}:
\begin{equation}
\begin{aligned}
& {\hat y_{ij}} = {u_i}^T{v_j}
\end{aligned}
\end{equation}

When optimizing model parameters, the pointwise loss function such as binary cross-entropy (BCE) and mean square error (MSE) etc. can be used \cite{he2017neural,hu2008collaborative,chen2020efficient}, but GCN generally use BPR loss due to its better performance. BPR loss chooses a user, an interactive item and a non-interactive item to form a triple, and the predicted value of interaction is expected to be greater than the predicted value of no interaction:
\begin{equation}
\begin{aligned}
& {L_{BPR}} =  - \sum\limits_{u = 1}^M {\sum\limits_{i \in {{\cal N}_u}} {\sum\limits_{j \notin {{\cal N}_u}} {\ln } } } \sigma \left( {{{\hat y}_{ui}} - {{\hat y}_{uj}}} \right) + \lambda {L_2}
\label{bpr_loss}
\end{aligned}
\end{equation}

Where $ \lambda $ is the parameter that controls the intensity of L2 regularization, and $ M $ is the total number of interactions. BPR loss is the key to achieving the final recommendation result, and we make it the main task of the SCL learning paradigm.

\subsection{BACKGROUND CONTRASTIVE LEARNING}
\label{cl}

\begin{figure*}
	\centering
		\includegraphics[scale=.25]{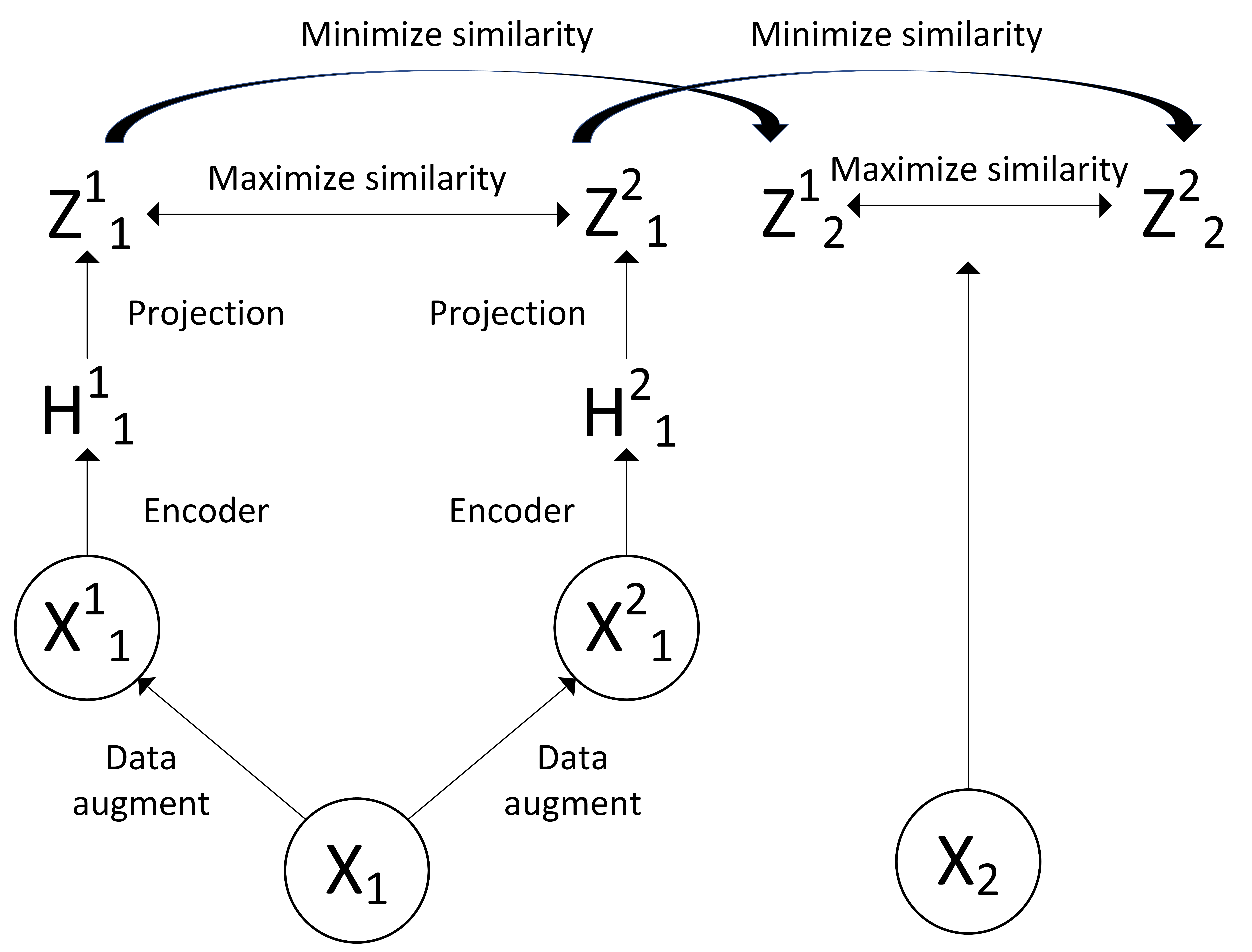}
	\caption{The framework of contrastive learning, it amit to minimize the distance of the augmented views generated by the same sample in the contrast loss space, while maximizing the distance of the augmented views generated by other samples in the same batch in that space.}
	\label{toy_example_cl}
\end{figure*}

The framework of contrastive learning is shown in the figure \ref{toy_example_cl}. It maximizes the similarity between augmented samples generated by the same data and minimizes the similarity with other samples in the same batch to learn representations. The framework includes four components: data augmentation, encoder, projection head, and contrast loss function \cite{chen2020simple}.

Specifically, the data augmentation module generates related views from the original data $ {X}_1 $ through some random data augmentation methods, denoted as $ {X^1}_1 $ and $ {X^2}_1 $. In the bipartite graph, the data augmentation methods we used include node drop, edge drop, and node replication. Different data augmentation methods have a great impact on the performance of comparative learning.

The encoder generates a representation vector $ {H^1}_1 $ and $ {H^2}_1 $ from the augmented view. In the user-item bipartite graph problem, the encoder is the GCN network.

The purpose of the projection head is to project the representation into the contrast loss space. The projection head is generally an MLP with a hidden layer. Many experiments on contrastive learning have proved that the projection head is beneficial to improve the performance of contrastive learning \cite{chen2020simple,chen2020improved,nguyen2021semi,verma2021towards}.

The contrast loss function which called InfoNCE aims to minimize the distance of the augmented views generated by the same sample in the contrast loss space, while maximizing the distance of the augmented views generated by other samples in the same batch in that space \cite{chen2020simple,sohn2016improved,wu2018unsupervised,oord2018representation}:
\begin{equation}
\begin{aligned}
& \ell_{i, j}=-\log \frac{\exp \left(\operatorname{sim}\left(\boldsymbol{z}_{i}, \boldsymbol{z}_{j}\right) / \tau\right)}{\sum_{k=1}^{2 N} 1_{[k \neq i]} \exp \left(\operatorname{sim}\left(\boldsymbol{z}_{i}, \boldsymbol{z}_{k}\right) / \tau\right)}
\label{infoNCE}
\end{aligned}
\end{equation}

Where $ \operatorname{sim}(\boldsymbol{u}, \boldsymbol{v})=\boldsymbol{u}^{\top} \boldsymbol{v} /\|\boldsymbol{u}\|\|\boldsymbol{v}\| $ denote the dot product between $\ell_{2}$ normalized $ \boldsymbol{u} $ and $ \boldsymbol{v} $. $ 1_{[k \neq i]} $ is an indicator function, when $ k $ is not equal to $ i $, the value is 1, otherwise it is 0. $ \tau $ denotes a temperature parameter.

\section{METHODOLOGY}
\label{method}

\begin{figure*}[htpb]
	\centering
		\includegraphics[scale=.20]{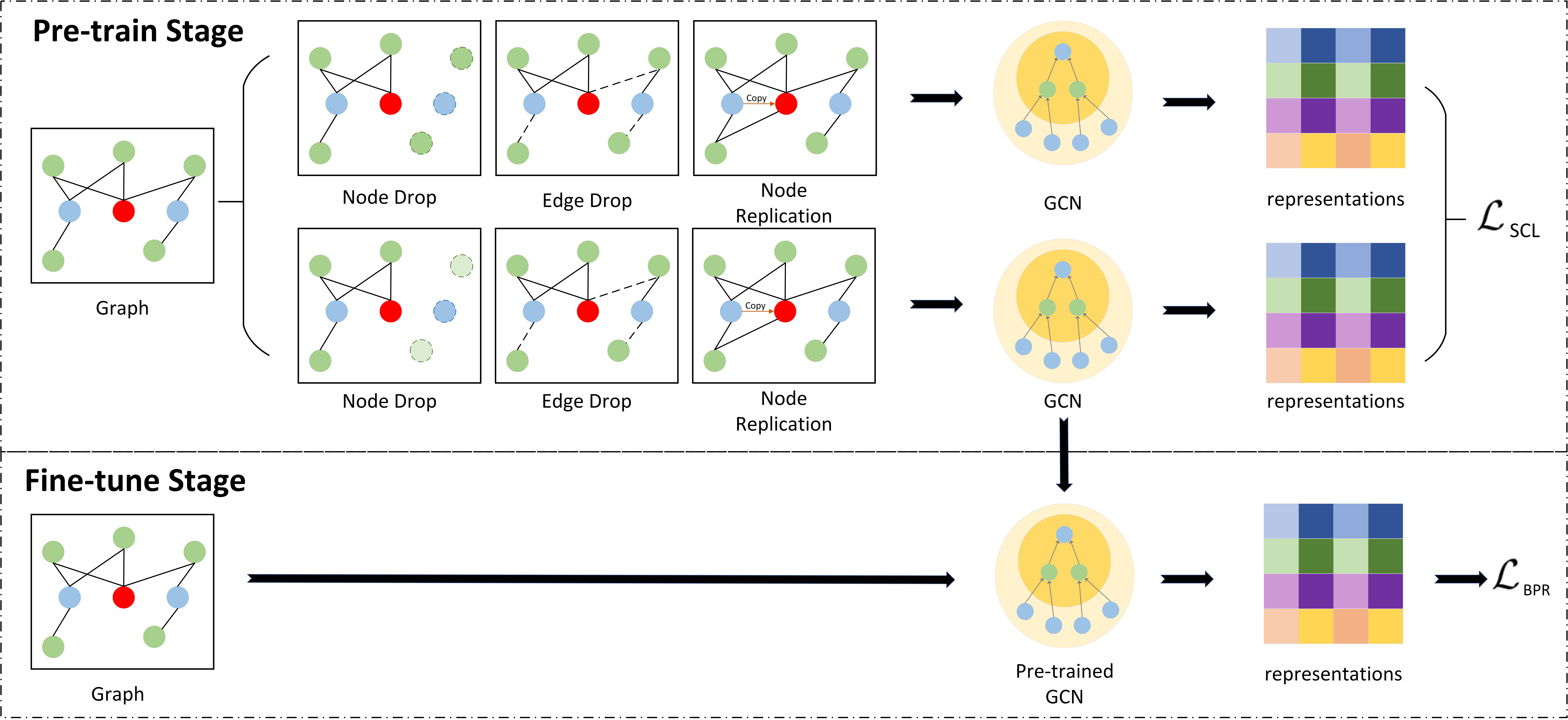}
	\caption{The framework of SCL, it employs multi-task learning to train the model with BPR loss as the main task and contrastive learning loss as the auxiliary task. Different from SGL, the contrastive loss of SCL is a supervised contrastive loss.}
	\label{framework}
\end{figure*}

We present the proposed paradigm of supervised constrastive learning for recommendation system. The learning process is divided into two parts, as shown in the Figure \ref{framework}. We use multi-task learning to train the model, with the BPR loss as the main task and contrastive learning loss as the auxiliary task \cite{rendle2012bpr}.In this section, we first introduce how to perform data augmentation to generate multiple views, and then introduce the design details of SCL.


\subsection{Data Augmentation}
\label{augmentation}

In SimCLR, the authors discusses various augmentation methods for images \cite{chen2020simple}, but for recommendation task, the data structure is different so that the data augmentation methods cannot be directly migrated. In the user-item bipartite graph recommendation task, the input data is a sparse matrix composed of interactions. We regard this data as the graph structure data as shown in the figure \ref{framework}. In order to learn the representation of each node fully, we need to design new data augmentation schemes for graph structure data. We detail the augmentation methods as follows:

\subsubsection{Node Drop(ND)}
\label{nd}

In SGL, the authors has designed various augmentation methods including node drop and edge drop for graph structure data \cite{wu2021self}. Node drop discards a certain node in the graph and any interactions(i.e. edge in graph) related to it according to probability $ {\rho _1} $:
\begin{equation}
\begin{aligned}
& V({\cal G}) = \left( {{{\bf{M _1}} } \odot {\cal N},{\cal E}} \right)
\label{for_nd}
\end{aligned}
\end{equation}

Where $ V({\cal G}) $ is the view generated by $ \cal G $ as input. $ \bf{M _1} $ is the masking vectors which applied on the node set, where 0 indicates that the corresponding node is discarded. $ \cal N $ represents the set of nodes, including user nodes set $ \cal U $ and item nodes set $ \cal V $. 

When performing data augmentation, we can generate multiple views by applying formula \ref{for_nd} multiple times on the original graph data. This augmentation can reduce the impact of high-frequency nodes on the representation by randomly drop some nodes, so that the representations can learn more from the long tail node.

\subsubsection{Edge Drop(ED)}
\label{ed}

Edge drop discards an certain interaction in the graph (i.e. edge in graph) according to the probability $ {\rho _2} $:
\begin{equation}
\begin{aligned}
& V({\cal G}) = \left( {{\cal N},{{\bf{M}}_2} \odot {\cal E}} \right)
\label{for_ed}
\end{aligned}
\end{equation}

Where $ \bf{M _2} $ is the masking vectors which applied on the edge set, where 0 indicates that the corresponding edge is discarded. $ \cal E $ represents the set of edges.

This augmentation randomly discards some edges, and it is expected that the representation learned by contrastive learning will not be affected by specific interactions, and the robustness of the representation will be improved.

\subsubsection{Node Replication(NR)}
\label{nr}

We believe that the above two augmentation methods only consider dropping a part of the information to alleviate the problem of noise and long-tail data, but do not consider increasing the diversity of the recommendation results by adding a part of the information. In order to further improve the diversity of the recommendation results and ensure that the representation of each node contains the user's possible interests as much as possible, we proposed a new data augmentation method called node replication.

Node replication will replace part of the interactions of the current node with the corresponding interactions of similar nodes according to probability $ {\rho _3} $. The similarity between the nodes is represented by the cosine, that is, the more similar is the interaction history, the more similar are the nodes. The matrix form of similarity calculation is as follows:
\begin{equation}
\begin{aligned}
& S = \frac{{{\cal G}{{\cal G}^T}}}{{\left| {\cal G} \right|}}
\end{aligned}
\end{equation}

Where $ {\cal G} $ is the sparse matrix represents the interaction history of each node. $ {\left| {\cal G} \right|} $ means row normalization of the matrix.Then for each node we will sort $ S $ and select the top $ N\% $ nodes as the similar nodes that can be used for replacement. It is worth noting that when calculating similarity, user-side nodes and item-side nodes are calculated separately. In addition, the similarity information obtained here will be further used in the calculation of contrastive learning in Section \ref{SCL}.

We divide the interaction history of each node into $ k $ segments, and randomly replace one of them when performing node replication according to the probability $ {\rho _3} $. Node replication is as follows:
\begin{equation}
\begin{aligned}
& V({\cal G}) = \left( {{{\bf{M}}_3} \odot {\cal N},{\cal E}} \right)
\end{aligned}
\end{equation}

Where $ \bf{M _3} $ is the masking vectors which apply on the node set, where 0 indicates that the corresponding node is applied node replication.

This augmentation randomly replaces part of the interactions, and it is expected that the representation can contain more information, increase the diversity of recommendation results, and improve the performance of the recommendation system.

\subsection{Supervised Comparative Learning}
\label{SCL}

After establishing multiple views of the node, we input them into the GCN respectively, and then perform contrastive learning loss on the generated representations. Different from InfoNCE in formula \ref{infoNCE}, which treats the views of the same node as a positive sample pair, and treats the views of any other different nodes as the negative samples, we propose a supervised contrastive learning loss, which treat similar nodes as the same class and all of those nodes are regarded as positive sample. We call this loss supervised InfoNCE(S-InfoNCE):
\begin{equation}
\begin{aligned}
& {\ell _{i,j}} =  - \log \frac{{\sum\limits_{k = 1}^{2N} {{1_{[j \in i]}}} \exp \left( {{\mathop{\rm sim}\nolimits} \left( {{{\bf{z}}_i},{{\bf{z}}_j}} \right)/\tau } \right)}}{{\sum\limits_{k = 1}^{2N} {{1_{[k \notin i]}}} \exp \left( {{\mathop{\rm sim}\nolimits} \left( {{{\bf{z}}_i},{{\bf{z}}_k}} \right)/\tau } \right)}}
\label{s_infonce}
\end{aligned}
\end{equation}

Where $ {{1_{[k \notin i]}}} $ denotes an indicator function, when $ k $ is not similar to $ i $, the value is 1, otherwise it is 0. While $ {{1_{[j \in i]}}} $ denotes that function, when $ j $ is similar to $ i $. $ {\mathop{\rm sim}\nolimits} \left( {i,j} \right) $ and $ \tau $ is same as infoNCE.

Our S-InfoNCE encourages the representations of similar nodes to be close to each other in the representation space to ensure their consistency. In addition, dissimilar samples are taken as negative samples to ensure that the representations of these nodes are significantly different. We take the recommendation task as the goal, and believe that similar users should have similar representations to facilitate collaborative filtering.

It is worth noting that the parameter $ \tau $ affects the performance of the representations learned by contrastive learning significantly. When $ \tau $ becomes larger, the samples of the same class will be more concentrated, the samples of different classes will be farther apart, and it will be more tolerant to difficult samples. When $ \tau $ becomes smaller, the representation of each sample will be more uniformly distributed on the hypersphere, and it will be more stringent for difficult samples. Our proposed S-InfoNCE treats similar nodes as a same class, and these similar nodes can be regarded as hard examples in InfoNCE. Since these nodes have been considered separately in S-InfoNCE, a relatively small $ \tau $ can be taken to obtain a more discriminative node representation and improve the performance of the recommender system.

\subsection{Multi-task Learning Loss of SCL}
\label{loss_SCL}

The learning paradigm of SCL apply multi-task learning strategy to jointly optimize the recommendation task loss(BPR loss, formula \ref{bpr_loss}) and the supervised contrastive loss(S-InfoNCE loss, formula \ref{s_infonce}):
\begin{equation}
\begin{aligned}
& {\cal L} = {{\cal L}_{BPR}} + {\lambda _1}{{\cal L}_{S - InfoNCE}} + {\lambda _2}{L^2}
\label{multi_task_learning}
\end{aligned}
\end{equation}

Where $ \lambda _1 $ is the hyperparameters to control the strengths of S-InfoNCE. $ \lambda _2 $ is the coefficient that controls the L2 regularization.

\subsection{Complexity Analyses of SCL}
\label{Complexity}

It is worth noting that our SCL learning paradigm will only perform back propagation with contrastive learning on the basis of GCN without introducing any additional parameters. In other words, the learning paradigm we proposed will not bring any parameter burden to the original model, and it will improve the recommendation performance while ensuring its high efficiency.

\section{EXPERIMENT}
\label{experiment}

We first describe experimental settings, and next compare with other state-of-the-art methods in Section \ref{performance}. To demonstrate the rationality and effectiveness of SCL, we conduct ablation study in Section \ref{ablation_study}. In addition, for parameters unique to SCL, we discuss them in Section \ref{study_scl}, separately.

\subsection{Experimental Settings}
\label{mfc}

To verify the effectiveness of the SCL for recommendation paradigm proposed in this paper, three public recommendation datasets including Gowalla \cite{liang2016modeling,he2016vbpr}, Yelp2018 \cite{wang2019neural,he2020lightgcn} and Amazon-Book \cite{he2016ups} are adopted for top-K recommendation and the results are compared with other state-of-the-art methods include NGCF, LightGCN and SGL. To reduce the experiment workload and keep the comparison fair, The dataset used is exactly the same as LightGCN paper used. The specific statistics of the three datasets are shown in the table \ref{datasets}.

\begin{table*}[htbp]
  \centering
  \caption{Statistics of the used three datasets}
    \begin{tabular}{c|c|c|c|c}
    \toprule
    \toprule
    \textbf{Dataset} & \textbf{Users\#} & \textbf{Items\#} & \textbf{Interactions\#} & \textbf{Density} \\
    \midrule
    \midrule
    \textbf{Gowalla} & 29858 & 40981 & 1027370 & 0.00\% \\
    \midrule
    \midrule
    \textbf{Yelp2018} & 31668 & 38048 & 1561406 & 0.0013 \\
    \midrule
    \midrule
    \textbf{Amazon-Book} & 52643 & 91599 & 2984108 & 0.00062 \\
    \bottomrule
    \bottomrule
    \end{tabular}%
  \label{datasets}%
\end{table*}%

\subsubsection{Experimental Metrics}
\label{metrics}

In top-k recommendation, we follow the strategy described in \cite{wang2019neural} and the ranking protocol described in \cite{wang2019neural}, NDCG (Normalized Discounted Cumulative Gain) and Recall are adopted to evaluate the recommendation performance. And consider the case where K to be 20 as LightGCN paper used.

\subsubsection{Implementation Details}
\label{details}

For fairness, all methods use the same training and testing data. The embedding size is fixed to 64 for all models and the embedding parameters are initialized with the Xavier method \cite{glorot2010understanding}. The Adam \cite{kingma2014adam} is used as optimizer to train the SCL, the batch size is set to 1024 and the learing rate is set to 0.001. The proposed SCL learning paradigm inherits the optimal parameters explored by LightGCN in the common part. For parameters unique to SCL, we conducted separate explorations. The $ \lambda _1 $ is searched in the range of $\left\{1 e^{-5}, 1 e^{-4}, 1 e^{-3},1 e^{-2},\right\}$, and the $ {\rho _3} $ is searched in the range of $\left\{0.05, 0.1 ,0.3,0.5,0.7\right\}$. The $ \tau $ is searched in the range of $\left\{0.05, 0.07 ,0.1,0.15,0.2\right\}$, and the $ N\% $ is searched in the range of $\left\{0.01\%, 0.1\% ,1\%,10\%,30\% \right\}$.
 
We implement multiple variants of SCL using different data augmentation methods, denoted by SCL-XX, where XX can be ND, ED, NR or a combination of them, representing node drop, edge drop, node replication, respectively.

\subsection{Performance Comparison}
\label{performance}

We compare SCL with LightGCN and NGCF on Gowalla, Yelp2018 and Amazon-Book, and the results are shown in the Table \ref{performance_compare}.

As can be seen from the table, our proposed SCL learning paradigm is more effective on the three datasets consistently, and has achieved significant performance improvements in two indicators. On Gowalla dataset, the improvement of SCL for LightGCN is not as large as that on Yelp2018 and Amazon-book. We believe this is due to the relatively small size of the Gowalla dataset, and LightGCN has learned relatively high-quality representations on less difficult datasets. In contrast, Yelp2018 and Amazon-Book datasets are larger and more difficult to train. For more difficult datasets, supervised contrastive learning as an auxiliary task effectively improves the quality of representations, making the improvement more obvious.

In addition, we also observe that SCL shows a consistent improvement with the number of model layers increases, the proportion of improvement is different of that. This is because increasing the number of layers only directly improves the representation quality of the GCN, while the existence of supervised contrastive learning in the SCL paradigm as an auxiliary task does not directly benefit from the increase in the number of model layers. In short, supervised contrastive learning is the same for different layers of models. When the quality of representation obtained by GCN is lower, SCL can achieve better improvement, as shown by the results of SCL on Yelp2018 and Amazon-Book.


\begin{table}[htbp]\scriptsize
  \centering
  \caption{Performance comparison with LightGCN and NGCF at different layers. The performance of LightGCN and NGCF are copied from its original paper. The percentage in brackets denote the relative performance improvement over LightGCN and The bold indicates the best result.}
  \resizebox{\textwidth}{!}{
    \begin{tabular}{c|c|c|c|c|c|c|c}
    \toprule
    \toprule
    \multicolumn{2}{c|}{\textbf{Database}} & \multicolumn{2}{c|}{\textbf{Gowalla}} & \multicolumn{2}{c|}{\textbf{Yelp2018}} & \multicolumn{2}{c}{\textbf{Amazon-Book}} \\
    \midrule
    \textbf{\#Layer} & \textbf{Method} & \multicolumn{1}{c|}{\textbf{Recall}} & \multicolumn{1}{c|}{\textbf{NDCG}} & \multicolumn{1}{c|}{\textbf{Recall}} & \multicolumn{1}{c|}{\textbf{NDCG}} & \multicolumn{1}{c|}{\textbf{Recall}} & \multicolumn{1}{c}{\textbf{NDCG}} \\
    \midrule
    \midrule
    \multirow{3}[2]{*}{\textbf{1Layer}} & \textbf{NGCF} & 0.1556 & 0.1315 & 0.0543 & 0.0442 & 0.0313 & 0.0241 \\
      & \textbf{LightGCN} & 0.1755 & 0.1492 & 0.0631 & 0.0515 & 0.0384 & 0.0298 \\
      & \textbf{SCL-NR} & \textbf{0.1764(+0.50\%)} & \textbf{0.1504(+0.80\%)} & \textbf{0.0655(+3.80\%)} & \textbf{0.0541(+5.05\%)} & \textbf{0.0432(+12.50\%)} & \textbf{0.0351(+17.79\%)} \\
    \midrule
    \multirow{3}[2]{*}{\textbf{2Layer}} & \textbf{NGCF} & 0.1547 & 0.1307 & 0.0566 & 0.0465 & 0.033 & 0.0254 \\
      & \textbf{LightGCN} & 0.1777 & 0.1524 & 0.0622 & 0.0504 & 0.0411 & 0.0315 \\
      & \textbf{SCL-NR} & \textbf{0.1790(+0.73\%)} & \textbf{0.1547(+1.50\%)} & \textbf{0.0673(+8.20\%)} & \textbf{0.0554(+9.92\%)} & \textbf{0.0448(+9.01\%)} & \textbf{0.0356(+13.02\%)} \\
    \midrule
    \multirow{3}[2]{*}{\textbf{3Layer}} & \textbf{NGCF} & 0.1569 & 0.1327 & 0.0579 & 0.0477 & 0.0337 & 0.0261 \\
      & \textbf{LightGCN} & 0.1823 & 0.1555 & 0.0639 & 0.0525 & 0.041 & 0.0318 \\
      & \textbf{SCL-NR} & \textbf{0.1830(+0.38\%)} & \textbf{0.1560(+3.20\%)} & \textbf{0.0683(+6.89\%)} & \textbf{0.0565(+7.62\%)} & \textbf{0.0459(+11.95\%)} & \textbf{0.0362(+13.84\%)} \\
    \bottomrule
    \end{tabular}}
  \label{performance_compare}%
\end{table}%

\begin{figure*}[htpb]
	\centering
		\includegraphics[scale=.25]{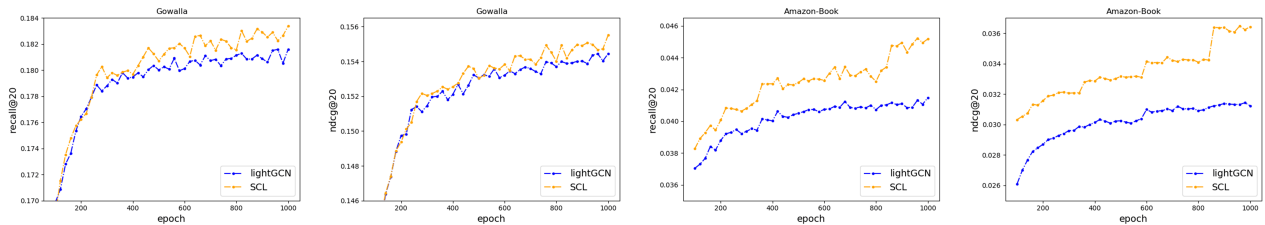}
	\caption{Training curves of LightGCN and SCL, which are evaluated by testing recall and testing ndcg per 20 epochs on Gowalla and Amazon-Book.}
	\label{curves}
\end{figure*}

The train curves of LightGCN and SCL which are evaluated by testing recall and testing ndcg per 20 epochs on Gowalla and Amazon-Book are shown as Figure \ref{curves}. Due to space limitations, we omit the results on the Yelp2018, which show the same trend.

As shown in the figure, we can observe that as the training process executes, SCL shows consistent advantages over LightGCN on both datasets. On the Gowalla dataset, SCL starts to outperform LightGCN on two metrics at 200 epoch, and the advantage gradually expands during the training process. Meanwhile, on the Amazon-Book dataset, SCL shows clear advantages over LightGCN from the beginning of training. This is because the Amazon-Book dataset is larger and more difficult to train. As an auxiliary task, SCL can help GCN converge faster and make the representations of nodes have higher quality. While the Gowalla dataset is smaller, GCN has learned relatively high-quality representations, and the improvement space of SCL is limited, making the advantage relatively small.

The above results further demonstrates the effectiveness of SCL, where supervised contrastive learning can effectively improve the quality of representations and improve the performance of recommender systems.

\subsection{Ablation Study}
\label{ablation_study}

\subsubsection{Effect of Augmentation method}
\label{effect6}

We conduct experiments on different data augmentation methods to illustrate their impact on performance. Table \ref{table_effect6} shows the results of different augmentation method. We ignore the results on Yelp2018 and Amazon-Book, which show the same trend as that on Gowalla. The following phenomena can be observed:
\begin{itemize}
\item When a single data augmentation method is used, ND, ED, and NR can all improve the performance. In contrast, NR performs better, which fully demonstrates the correctness and effectiveness of our motivation for proposing NR. Furthermore, we also observe that NR is more computationally efficient than ND and ED (about twice that of ND and three times that of ED on the Gawalla dataset), which is beneficial for the practical application of NR.
\item Combining NR with different data augmentation methods has limited effect improvement. This means that NR has generated enough information for SCL to learn. Using NR alone as a data augmentation method is sufficient for SCL to learn node representations for recommendation results.
\end{itemize}


\begin{table}[htbp]
  \centering
  \caption{The performance of different augmentation methods}
    \begin{tabular}{c|c|c}
    \toprule
    \toprule
    \textbf{Database} & \multicolumn{2}{c|}{\textbf{Gowalla}} \\
    \midrule
    \textbf{Method} & \textbf{Recall} & \textbf{NDCG} \\
    \midrule
    \midrule
    \textbf{SCL-ND} & 0.1818 & 0.1548 \\
    \textbf{SCL-ED} & 0.1826 & 0.155 \\
    \textbf{SCL-NR} & \textbf{0.183} & \textbf{0.1553} \\
    \textbf{SCL-ND+NR} & 0.1829 & \textbf{0.1553} \\
    \textbf{SCL-ED+NR} & 0.1828 & 0.1552 \\
    \textbf{SCL-ND+ED+NR} & 0.1829 & \textbf{0.1553} \\
    \bottomrule
    \end{tabular}%
  \label{table_effect6}%
\end{table}%

\subsubsection{Effect of Contrastive Learning}
\label{effect7}

To illustrate the advantages of supervised contrastive learning over unsupervised contrastive learning. We compare the performance of SCL and SGL on Gowalla and Amazon-book, and the results are shown in the Table \ref{table_effect7}.

It can be found that SCL shows consistent advantages on both datasets. Among them, the advantage is more significant on the Yelp2018 dataset. As mentioned in Section \ref{performance}, Gowalla is relatively small in scale and is easier to train, which leaves limited space for improvement in contrastive learning. On larger datasets, SCL can show more obvious advantages.

In short, the results in Table \ref{table_effect7} demonstrate that SCL can help recommender systems learn higher-quality representations. Due to the introduction of supervised information, the model is guided to adjust the distribution of nodes in the representation space, so that the representations of similar nodes are close to each other in the representation space, thereby improving the performance of the recommender system.

\begin{table*}[htbp]
  \centering
  \caption{Performance comparison with SGL. The performance of SGL on Yelp2018 and Amazon-book are copied from its original paper and the performance on Gowalla is be realized in this paper. The bold indicates the best result.}
    \begin{tabular}{c|c|c|c|c}
    \toprule
    \toprule
    \textbf{Database} & \multicolumn{2}{c|}{\textbf{Gowalla}} & \multicolumn{2}{c|}{\textbf{Yelp2018}} \\
    \midrule
    \textbf{Method} & \textbf{Recall} & \textbf{NDCG} & \textbf{Recall} & \textbf{NDCG} \\
    \midrule
    \midrule
    \textbf{SGL} & 0.1827 & 0.1558 & 0.0675 & 0.0555 \\
    \textbf{SCL-NR} & \textbf{0.183} & \textbf{0.156} & \textbf{0.0683} & \textbf{0.0565} \\
    \bottomrule
    \end{tabular}%
  \label{table_effect7}%
\end{table*}%

\subsection{Study of SCL}
\label{study_scl}
To further illustrate the impact of unique parameters of SCL on recommendation performance, we explore each parameter. In addition, we also try to use SCL as the pre-training part of GCN. Due to the space limitation, we omit the results on Amazon-Book and Yelp2018 which have same trends to that on Gowalla dataset.

\subsubsection{Effect of $\lambda _1$}
\label{effect1}

\begin{figure}[htpb]
	\centering
		\includegraphics[scale=.45]{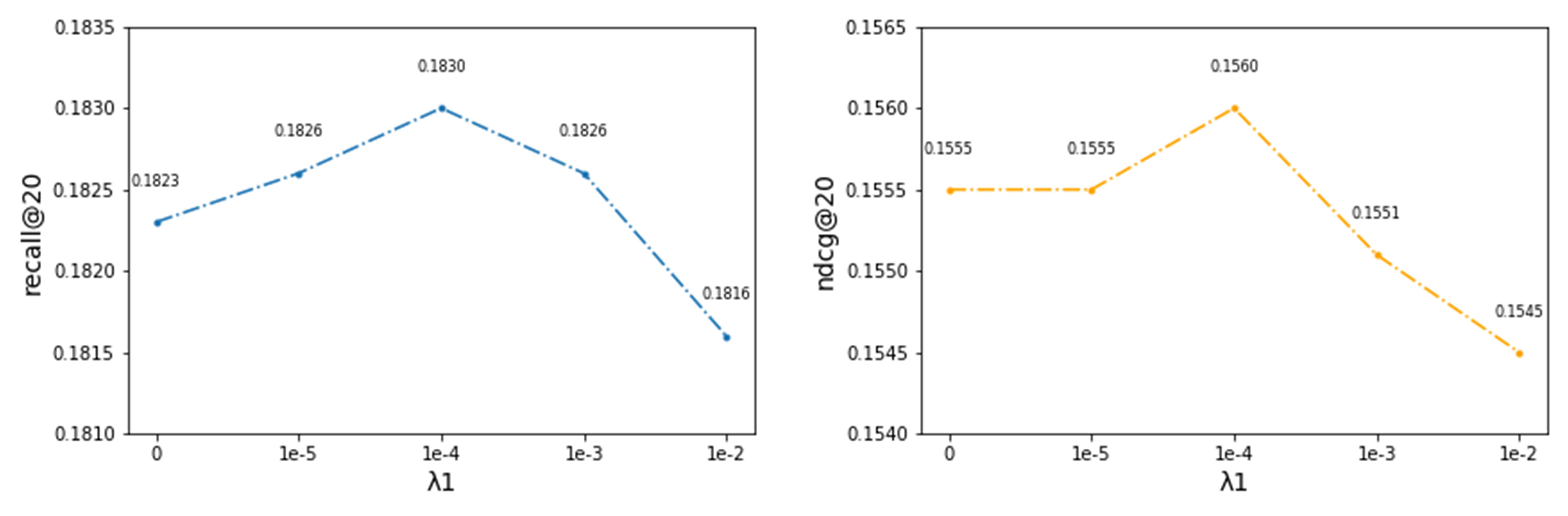}
	\caption{The trend of the performance of SCL as a function of $ \lambda _1 $.}
	\label{effect1_pic}
\end{figure}

The role of parameter $ \lambda _1 $ is to control the importance of contrastive learning in multi-task learning. When $ \lambda _1 $ is 0, multi-task learning degenerates to single-task learning which use BPR loss only. Figure \ref{effect1_pic} shows the trend of SCL performance as a function of parameter $ \lambda _1 $, we can observe the following two phenomena:

\begin{itemize}
\item As $ \lambda _1 $ increases from 0 to 1e-4, the performance of SCL shows consistent improvement on Recall@20 and NDCG@20. This shows that contrastive learning as an auxiliary task effectively improves the representation quality of nodes, which in turn improves the performance of the recommender system.
\item With the further increase of $ \lambda _1 $, the performance of SCL decreases. This means that excessive weighting of contrastive learning as an auxiliary task may affect the learning of the BPR main task and negatively affect performance. Therefore, we recommend carefully adjusting $ \lambda _1 $ when using it.
\end{itemize}

\subsubsection{Effect of ${\rho _3}$}
\label{effect2}

\begin{figure}[htpb]
	\centering
		\includegraphics[scale=.45]{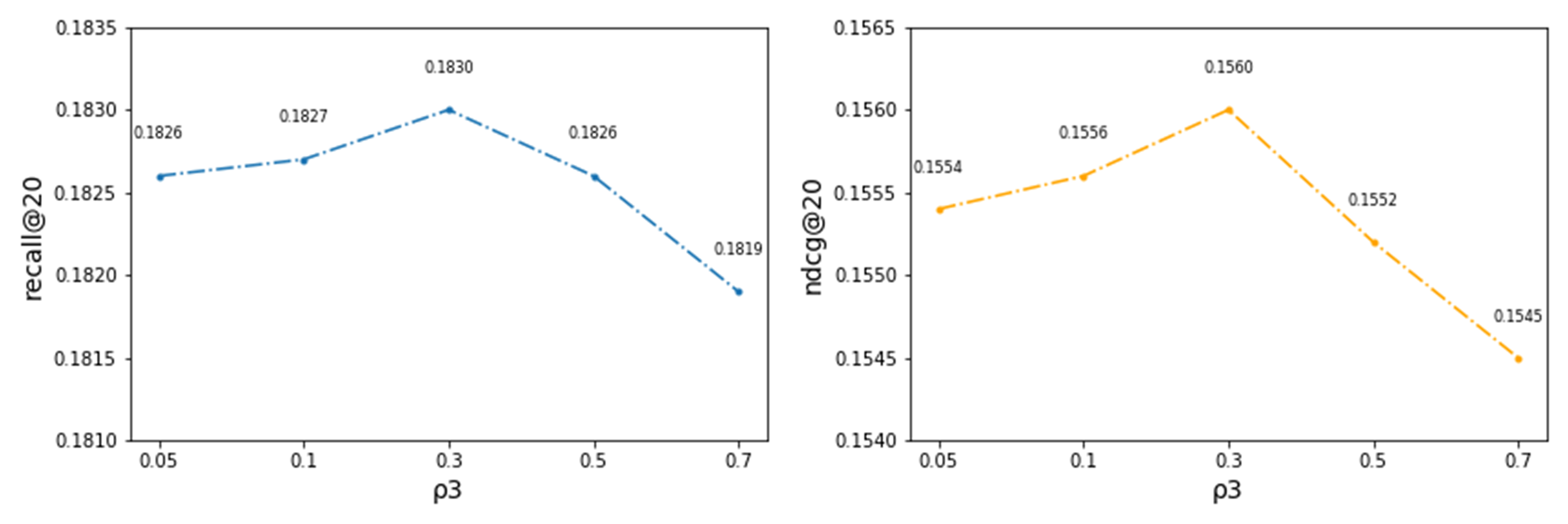}
	\caption{The trend of the performance of SCL as a function of ${\rho _3}$.}
	\label{effect2_pic}
\end{figure}

The parameter ${\rho _3}$ controls the probability of performing node replication during data augmentation. We fix $ \lambda _1 $ as 1e-4 and observes the performance of SCL under different ${\rho _3}$. The results are shown in Figure \ref{effect2_pic}. we can find out that:

\begin{itemize}
\item As ${\rho _3}$ increases from 0.05 to 0.3, the performance of SCL shows consistent improvement on Recall@20 and NDCG@20. This means that increasing the learning difficulty of the contrastive learning task to a certain extent helps to improve the quality of node representations.
\item With the further increase of ${\rho _3}$, the performance of SCL decreases. This means that an excessively large ${\rho _3}$ will not only bring additional computational burden, but also make it difficult for the contrastive learning loss to converge.
\end{itemize}

\subsubsection{Effect of $N\%$}
\label{effect3}

\begin{figure}[htpb]
	\centering
		\includegraphics[scale=.45]{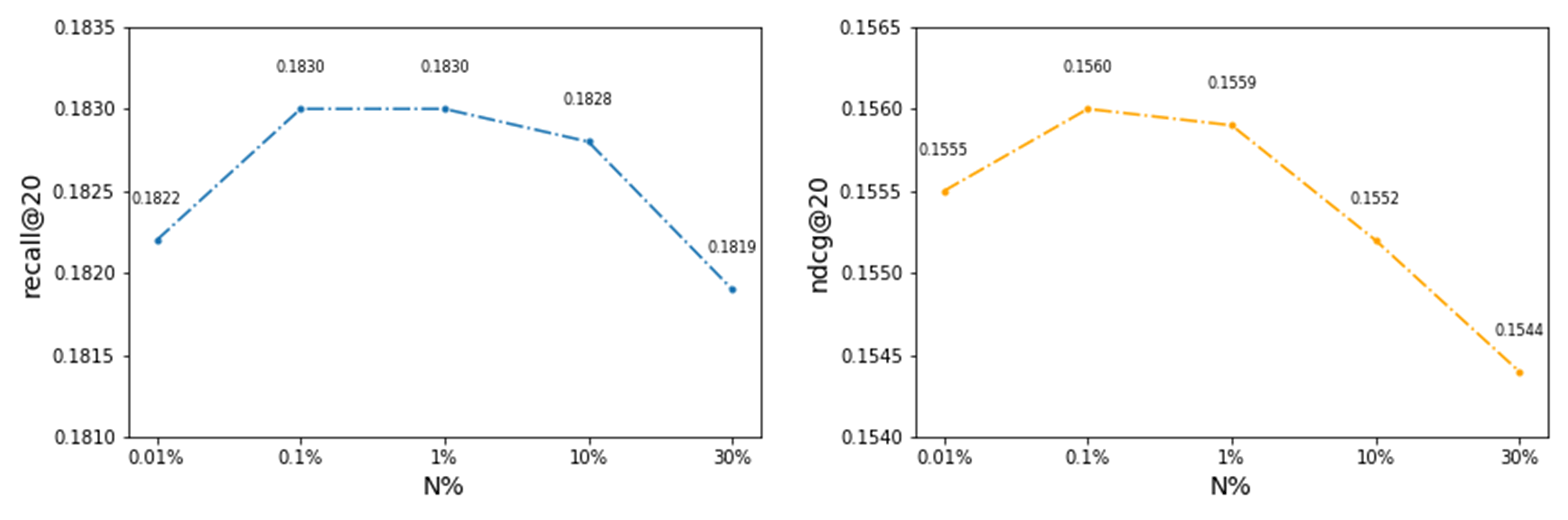}
	\caption{The trend of the performance of SCL as a function of $N\%$.}
	\label{effect3_pic}
\end{figure}

The parameter $N\%$ controls number of similar samples. We fix $ \lambda _1 $ and ${\rho _3} $ as 1e-4, 0.3 respectively, and observes the performance of SCL under different $N\%$. The results are shown in Figure \ref{effect3_pic}. we can find out that the performance of SCL shows the characteristics of first increase and then decrease. A too large $N\%$ will make the samples with lower similarity be regarded as positive samples, and a too small $N\%$ will make SCL degenerate into ordinary unsupervised contrastive learning.

Different datasets may need different $N\%$. We suggest carefully adjusting the setting of $N\%$ in the application of SCL. A too large or too small $N\%$ will have a significant impact on the performance of the recommendation system.

\subsubsection{Effect of $\tau$}
\label{effect4}

\begin{figure}[htpb]
	\centering
		\includegraphics[scale=.45]{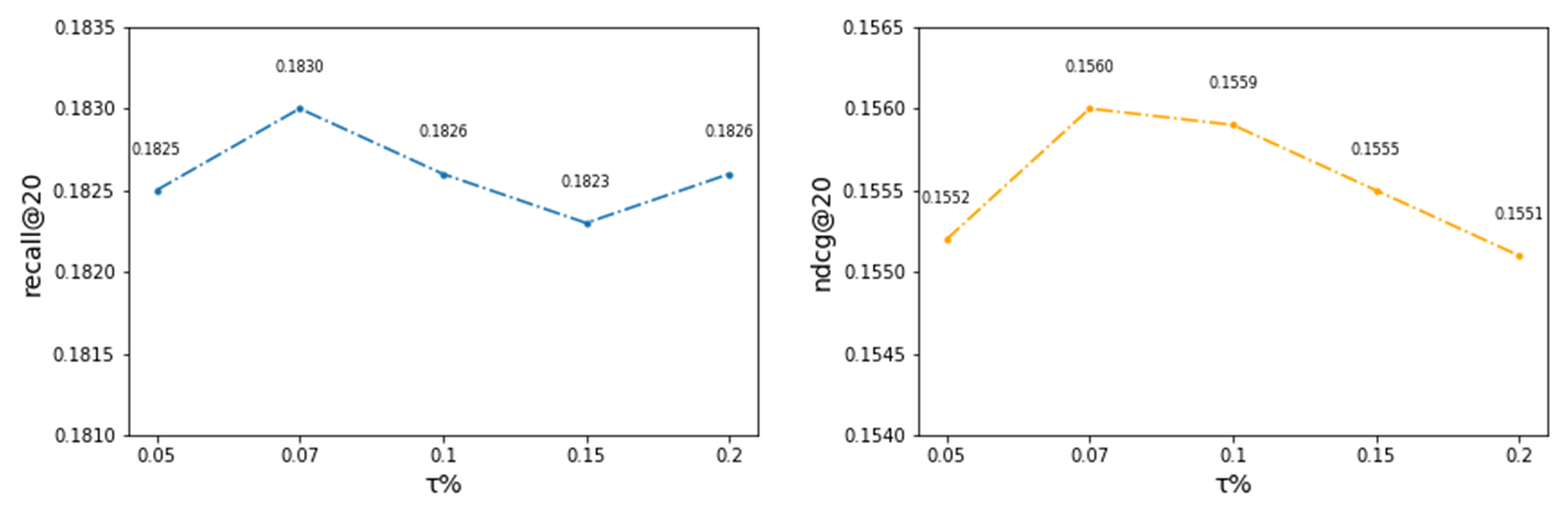}
	\caption{The trend of the performance of SCL as a function of $\tau$.}
	\label{effect4_pic}
\end{figure}

The parameter $\tau$ controls the uniformity of the distribution of the representations on the hypersphere. The trend of performance of SCL with $\tau$ is shown in the Figure \ref{effect4_pic}. We can observe that:

\begin{itemize}
\item As mentioned earlier, SCL considers the effect of hard negative samples in the S-InfoNCE, so a smaller $\tau$ can be used to achieve better results. However, if $\tau$ is too small, the distance between the representations of similar nodes will become farther, which is contrary to the optimization goal of the recommendation system and will degrade the performance of the recommender system.
\item With the continuous increase of $\tau$, the representation of nodes will be smoother and the discrimination will be reduced, which makes dissimilar samples closer in the representation space, so that the recommendation system may recommend items that have low correlation or are not related to users. In result, the performance of the recommendation system being reduced.
\end{itemize}

\subsubsection{Effect of Pre-training}
\label{effect5}

We compare the difference between applying SCL in a pre-training way and applying SCL in a multi-task learning way. During pre-training, we set the epoch to 1000, the learning rate to 0.01, the batch size to 1024, the optimization method is Adma, and the L2 regularization coefficient is set to 1e-4. All settings in the fine-tuning stage are consistent with LightGCN. The results are shown in Table \ref{effect5_table}. 

We find that $ {\rm{SCL}}{{\rm{ - }}_{{\rm{pre - train}}}} $ has a consistent decrease in Recall and ndcg compared to $ {\rm{SCL}}{{\rm{ - }}_{\rm{multi - task}}} $, which means that pre-trained representations may be unfavorable for recommendation task learning. This is because the loss of contrastive learning only guides the distribution of representation in representation space, but does not have interactive information. This may make the representations obtained by $ {\rm{SCL}}{{\rm{ - }}_{{\rm{pre - train}}}} $ contrary to the recommended task. To further improve the effect of $ {\rm{SCL}}{{\rm{ - }}_{{\rm{pre - train}}}} $, it may be necessary to warm up or increase the learning rate in the fine-tuning stage to make the model jump out of the local optimization, which will further increase the computational cost of the model.

Contrastive learning can help improve the performance of recommender systems only when the recommendation task is taken as the main task, this is because comparative learning, as an auxiliary task, can help the model improve the quality of representations of nodes, and will not affect the learning of BPR loss. therefore, we propose to apply SCL in a multi-task learning way.

\begin{table}[htbp]
  \centering
  \caption{The difference between $ {\rm{SCL}}{{\rm{ - }}_{{\rm{pre - train}}}} $ and $ {\rm{SCL}}{{\rm{ - }}_{\rm{multi - task}}} $}
    \begin{tabular}{c|c|c}
    \toprule
    \toprule
    \textbf{Database} & \multicolumn{2}{c}{\textbf{Gowalla}} \\
    \midrule
    \textbf{Method} & \textbf{Recall} & \textbf{NDCG} \\
    \midrule
    \midrule
    \textbf{$ {\rm{SCL}}{{\rm{ - }}_{{\rm{pre - train}}}} $} & \textbf{0.183} & \textbf{0.156} \\
    \textbf{$ {\rm{SCL}}{{\rm{ - }}_{\rm{multi - task}}} $} & 0.1782 & 0.1522 \\
    \bottomrule
    \end{tabular}%
  \label{effect5_table}%
\end{table}%

\section{Conclusion}
\label{conclu}

In this work, we recognize the pitfalls of direct application of contrastive learning to the recommendation system domain, and propose a supervised contrastive learning paradigm for recommender systems. We use the interaction information of the user-item bipartite graph to construct supervised information, and improve the contrastive learning loss, so that similar nodes in a batch are close to each other and dissimilar nodes are farther apart in the representation space. Furthermore, to improve the diversity of recommendation results, we also propose a new data augmentation method called node replacement. We conduct extensive comparative trials and ablation experiments on three benchmark datasets, and the experimental results demonstrate the superiority of SCL.

In this work, supervised information is currently used to improve the contrastive learning loss only. In future work, we hope to make fully use of supervised information. For example, using supervised information to mine hard samples or improve BPR loss will be very interesting research directions.

\bibliographystyle{unsrt}  
\bibliography{references}

\end{document}